\documentclass[showpacs,prb,reprint,preprintnumbers,amsmath,amssymb,superscriptaddress,floatfix]{revtex4-1}
\usepackage{graphicx}
\usepackage{dcolumn}

\begin{document}
\newcommand{\oH}{\ensuremath{\mathcal{H}}}
\newcommand{\oI}{\ensuremath{\mathcal{I}}}
\newcommand{\oS}{\ensuremath{\mathcal{S}}}
\newcommand{\oQ}{\ensuremath{\mathcal{Q}}}

\title{Quadrupolar spectra of nuclear spins in strained InGaAs quantum dots}

\author{Ceyhun \surname{Bulutay}}
\email{bulutay@fen.bilkent.edu.tr}
\affiliation{Department of Physics, Bilkent University, Ankara 06800, Turkey}
\affiliation{Institute of Quantum Electronics, ETH-Z\"{u}rich, CH-8093 Z\"{u}rich, Switzerland}
\date{\today}

\begin{abstract}
Self-assembled quantum dots (QDs) are born out of lattice mismatched ingredients 
where strain plays an indispensable role. Through the electric quadrupolar coupling, 
strain affects the magnetic environment as seen by the nuclear spins. 
To guide prospective single-QD nuclear magnetic resonance (NMR) as well as 
dynamic nuclear spin polarization experiments, an atomistic insight to the strain and 
quadrupolar field distributions is presented. A number of implications of the 
structural and compositional profile of the QD have been identified. 
A high aspect ratio of the QD geometry enhances the quadrupolar interaction. The inclined 
interfaces introduce biaxiality and the tilting of the major quadrupolar principal 
axis away from the growth axis; the alloy mixing of gallium into the QD enhances both of
these features while reducing the quadrupolar energy.
Regarding the NMR spectra, both Faraday and Voigt geometries are investigated, unraveling
in the first place the extend of inhomogeneous broadening and the appearance of the 
normally-forbidden transitions. Moreover, it is shown that from the main extend of the 
NMR spectra the alloy mole fraction of a single QD can be inferred.
By means of the element-resolved NMR
intensities it is found that In nuclei has a factor of five dominance over those of As.
In the presence of an external magnetic field, the borderlines between the quadrupolar and Zeeman 
regimes are extracted as 1.5~T 
for In and 1.1~T for As nuclei. At these values the nuclear spin depolarization rates
of the respective nuclei get maximized due 
to the noncollinear secular hyperfine interaction with a resident electron in the QD.
\end{abstract}


\maketitle

\section{Introduction}
In the coherent control of electron spins within a solid state environment, such as 
in quantum dots (QDs), the nuclear spin reservoir acts as the main source of 
decoherence.\cite{khaetskii02,merkulov02,desousa03} Recent studies now assure that 
nuclear spin bath can be tamed so as to counter the detrimental effect it may have on the carrier 
spins.\cite{kane98,taylor03,petta05,fischer9,bluhm11} Therefore, in the emerging state of 
understanding, an alternative is to utilize nuclear spins as a resource,\cite{ardavan11} for instance as 
a natural qubit memory (for an extended review and references, see, Ref.~\onlinecite{liu10}).
In these studies, self-assembled QDs have been one of the archetypal systems. 
Within the full gamut of possible combinations of self-organized materials,\cite{stangl04} 
the InAs on GaAs system stands out due to extensive research efforts 
devoted over more than a decade to their growth and optoelectronic 
characterizations.\cite{petroff03} For the purposes of controlling 
the spin dynamics in InAs QDs, several 
groups\cite{lai06,eble06,tart07,urbaszek07,makhonin08,kroner08,xu09,latta09,
cherbunin09,makhonin09,eble09,krebs10,cherbunin11} have studied various aspects of the 
optical orientation\cite{meier84} of an electron spin and its transfer to the nuclear spins
via the hyperfine interaction leading to the dynamic nuclear spin polarization.\cite{lampel68} 

One of the remarkable achievements that paved the way to the coherent manipulation of nuclear spin 
dynamics has been the nuclear magnetic resonance (NMR) of a {\em single} QD.\cite{gammon97,gammon01} 
With further improved precision, position selective control of small groups of nuclear spins
inside a dot is being pursued.\cite{makhonin10} This so-called optically detected NMR
is initiated by the circularly polarized excitation of a spin-polarized electron spin 
inside a QD that polarizes the nuclear spins through the hyperfine interaction. The 
Overhauser field established by the polarized nuclei acts back on the electronic system, 
which can be externally measured over an excitonic Zeeman splitting. Additionally if an rf 
magnetic field resonant with nuclear spin transitions is incident, it depolarizes some 
of the nuclear spins reducing the Overhauser field, which can in turn be detected from a shift
in the optical emission spectra. Along this line, a recent demonstration utilized
a sequence of two phase-locked rf pulses to induce coherent rotations of a targeted group of
nuclear spins optically pumped to a high polarization degree.\cite{makhonin11} The advantage
of this technique is that it enables full coherent control over the Bloch sphere and yet on the
order of microsecond time scales. 

These NMR experiments were performed on GaAs interface-fluctuation QDs 
with the deliberate aim of avoiding any strain to keep the resonances 
narrow.\cite{makhonin10,makhonin11} On the other hand strain is an integral part of 
self-assembled InAs QDs.\cite{petroff03}
The existing negative sentiments for the strain in the context of electron and nuclear spin dynamics
have been dramatically reversed by the work of Dzhioev and Korenev.\cite{dzhioev07}
Actually, several decades ago it was experimentally shown that anisotropic 
strain in a III-V crystal 
lattice causes local electric field gradients (EFG) with which a spin-$I$ nucleus 
with $I\ge 2$ interacts because of its 
quadrupolar moment.\cite{sundfors69a,sundfors69b,sundfors74}
This quadrupolar interaction (QI) splits the nuclear spin degeneracy even in the 
absence of an external magnetic field,\cite{abragam61,slichter78} 
hence it energetically suppresses the nuclear spin flip events, stabilizing 
the electron spin orientation confined in the QD.\cite{dzhioev07} 
With this paradigm shift, strain is no longer just a nuisance but something to be exploited 
as a new degree of freedom to tailor the magnetic environment of InAs QDs.

The aim of this work is to offer an atomistic understanding of the interesting physics 
arising from the coexistence of QI together with the dc and rf magnetic fields. This is in accord with
the current progress of the experimental techniques having the goal to address and control
relatively few number of nuclei within a QD.\cite{makhonin10,makhonin11} Starting from the
behavior of the spatial variation of the strain tensor, we trace the factors 
that affect QI, and identify primarily the high aspect ratio of the QD as a factor for 
enhancing it. 
Considering different magnetic field orientations and the effect of random alloy mixing 
i.e., In$_x$Ga$_{1-x}$As QDs, critical insight is gained on the vulnerability of each 
nuclear spin transition in the NMR spectrum under inhomogeneous broadening.
An important complication is that the mole fraction of the constituents can vary 
appreciably from dot to dot within the same sample,\cite{biasol11,koenraad-chapter}
rendering the in situ compositional identification of a targeted QD so far not practical.
We demonstrate that this can be readily extracted from the main span of the NMR spectra.
Moreover, our element-resolved spectra present 
crucial information for the labeling of the features in the rather complicated overall spectra. 
For the typical QD parameters reported in relevant experiments, we extract an effective quadrupolar 
field, $B_Q$ of 1.5~T (1.1~T) for the In (As) nuclei, as a borderline below which QI
dominates. The atomistic picture that we acquire also provides us the distribution of the 
quadrupolar principal axes within the QD. We make use of this information in working out 
the nuclear spin depolarization due to noncollinear secular hyperfine interaction 
arising from the tilting of 
the major quadrupolar axis from the optical axis. The depolarization time drops to a minimum
on the order of an hour at the magnetic field coinciding with $B_Q$, as a manifestation of
the strong competition between the QI and the Zeeman field at this value.

Regarding the organization of the paper: in Sec.~II we give the theoretical details for 
the strain, QI, and the nuclear spin depolarization; in Sec.~III some information about our 
QD structures is given, followed by the results grouped into
strain and quadrupolar splittings, NMR spectra, and the noncollinear secular hyperfine interaction;
in Sec.~IV we summarize our conclusions, and append a section on the matrix elements 
for obtaining the energy spectra and the rf-initiated nuclear spin transitions under an arbitrary EFG.

\section{Theory}
Although the theoretical ingredients employed in this work are not new, 
an account of the procedures and the mathematical models will be helpful for clarifying how the 
results are obtained.

\subsection{Strain}
From a computational point of view, to identify the strain profile, the ionic 
relaxation of the QD and the host matrix atoms 
is needed. The technique used for this purpose is molecular statics as implemented in the LAMMPS 
code.\cite{lammps} Here, the main input is the semi-classical force field 
that governs the atomic interactions. The preferred choice for group~IV and III-V semiconductors 
are the Abell-Tersoff potentials.\cite{abell-tersoff} We make use of the recent parametrization 
by Powell {\em et al.} who fitted their force field expression to a large set of cohesive and elastic 
properties obtained 
from density functional theory.\cite{powell07} With these tools and for a chosen compositional 
profile, the relaxed atomic positions become readily available.\cite{ramsey08}
Next, one needs to extract the strain state, again in an atomistic level as we aim for the
EFGs at each nuclear site. Among the several possible strain measures, we adopt the one proposed by 
Pryor {\em et al} because of its inherent compatibility with tetrahedral bonds, as in the present 
case.\cite{pryor98} In this measure, the critical task is to correctly align an unstrained 
reference tetrahedron for each local bonding topology. This local strain attains a 
very rapid variation especially for the random alloy mixing of 
In$_x$Ga$_{1-x}$As. Therefore, like Bester {\em et al.},\cite{bester06} we average the strain tensor;
in our case over the twelve neighboring same-ionicity sites, which we also apply to the pure 
InAs QDs as well.

\subsection{Quadrupolar interaction}
In the linear elastic limit, we can express the EFG tensor components $V_{ij}$
in any orthogonal coordinate frame\cite{note1} using the computed local strain 
tensor $\epsilon_{ij}$ 
as:
\begin{equation}
V_{ij}\equiv\frac{\partial^2V}{\partial x_i\partial x_j}=\sum_{k,l=1}^3 S_{ijkl}\epsilon_{kl},
\end{equation}
where $S$ is the fourth-rank gradient elastic tensor.\cite{shulman57,note2}
Transforming this tensor expression to Voigt notation in the cubic crystallographic frame we get
\begin{equation}
V_{\mu}=\sum_{\nu=1}^6 S_{\mu\nu}\epsilon_{\nu}\, .
\end{equation}
As the trace of the EFG is unobservable\cite{harrison62} it is conveniently set to zero, 
$\sum_{i}V_{ii}\to 0$, which leads to $S_{11}=-2S_{12}$. This
results in the following explicit relations for the EFG tensor components in the mixed
Voigt and tensor notation
\begin{eqnarray}
V_{zz} & = & S_{11}\left[ \epsilon_{zz}-\frac{1}{2}\left( \epsilon_{xx}+\epsilon_{yy}\right)\right]\, , \\
V_{xy} & = & V_{yx}=2 S_{44}\epsilon_{xy}\, ,
\end{eqnarray}
with the remaining components being obtained by their cyclic permutations.\cite{warn}
The EFG tensor couples to the nuclear quadrupole moment tensor (operator) $\oQ_{\alpha\beta}$ 
through the Hamiltonian\cite{slichter78}
\begin{eqnarray}
\oH_Q & = &\frac{1}{6}\sum_{\alpha,\beta}V_{\alpha\beta}\oQ_{\alpha\beta}\, \nonumber ,\\
& = & \frac{eQ}{6I(2I-1)}\sum_{\alpha,\beta}V_{\alpha\beta}\left[\frac{3}{2} \left(
\oI_\alpha\oI_\beta+\oI_\beta\oI_\alpha\right)-
\delta_{\alpha\beta}\oI^2\right]\, \nonumber,
\end{eqnarray}
where $\mathbf{\oI}$ is the dimensionless spin operator, $e$ is the electronic 
charge, $I$ is 9/2 for In, and 3/2 for As and Ga nuclei,
and $Q$ is the electric quadrupole moment of the nucleus.
This expression gets simplified in the frame of EFG principal axes ($V_{IJ}\equiv 0$, for $I\ne J$) as
\begin{equation}
\oH_Q=\frac{e^2qQ}{4I(2I-1)}\left[ 3\oI^2_Z-\oI^2+\eta\frac{\oI^2_+-\oI^2_-}{2} \right]\, ,
\end{equation}
where $\oI_\pm\equiv \oI_X\pm i\oI_Y$, $q\equiv V_{ZZ}/e$ is the field gradient parameter, 
and $\eta=(V_{XX}-V_{YY})/V_{ZZ}$ is the biaxiality (asymmetry) parameter. 
The former is the primary coupling constant of QI, and the 
latter determines the mixing between the free nuclear spin magnetic quantum numbers.
In the most general case that we shall consider, in addition to quadrupolar part $\oH_Q$, 
a nuclear spin will have interactions with dc and rf magnetic fields in the form
$\oH_M=-\gamma\hbar\mathbf{\oI}\cdot\mathbf{{B_0}}$ and 
$\oH_{\hbox{\small rf}}
=-\gamma\hbar\mathbf{\oI}\cdot\mathbf{B^{\hbox{\small rf}}}
\cos\omega_{\hbox{\small rf}} t$, where $\gamma$ is the 
gyromagnetic ratio of the nucleus. The weak rf part can safely be treated perturbatively,
whereas the dc magnetic field can be strong and gives rise to Zeeman effect; 
we shall abbreviate the stationary states under both quadrupolar and Zeeman splittings as 
the QZ spectra.\cite{noteQZ} Individual matrix elements for obtaining the QZ spectra for an arbitrary
EFG and $\mathbf{B}_0$ as well as the 
$\oH_{\hbox{\small rf}}$-initiated 
transition rates are given in the Appendix section for the sake of completeness.

\subsection{Noncollinear secular hyperfine interaction}
Quite commonly in the nuclear spin experiments there exists an electron in the QD 
with an optically oriented spin along the growth 
axis.\cite{lai06,eble06,tart07,urbaszek07,makhonin08,kroner08,xu09,latta09,
cherbunin09,makhonin09,eble09,krebs10,cherbunin11} Through the hyperfine 
interaction it then polarizes the QD nuclear spins, termed as the dynamic 
nuclear polarization.\cite{lampel68} 
Deng and Hu have suggested that in the presence of quadrupolar mixing a new spin
depolarization channel becomes possible through the hyperfine coupling.\cite{deng05}
If we leave out the weaker anisotropic dipolar part
in the $s$-type conduction band of III-V semiconductors,\cite{coish10} the hyperfine 
interaction can be represented by the isotropic Fermi contact term\cite{huang10}
\begin{equation}
\oH_{\hbox{\small hf}}=\sqrt{f_e}
A_{\hbox{\small hf}}|\psi(\mathbf{R})|^2 \left[ 
\underbrace{\oI_z\oS_z}_{\hbox{\small secular}}
+\frac{1}{2}\underbrace{\left(\oI_+\oS_- + \oI_-
\oS_+\right)}_{\hbox{\small non-secular}}
\right],
\end{equation}
where $\oS_z,\oS_\pm$ are the $z$-component and the raising/lowering 
electron spin operators, 
$A_{\hbox{\small hf}}$ is the hyperfine constant 
of the nucleus, $f_e$ is the average fraction of 
the time electron is inside the QD (i.e., hyperfine interaction is on) for which we use 
the value 0.035.\cite{maletinsky07,huang10}  At a nuclear site, $\mathbf{R}$,
$\psi(\mathbf{R})$ is the electron wave function, that we
simply approximate with a $z$-varying (height-dependent) Gaussian profile over the QD as 
$\psi(\rho,z)\propto e^{-\left(3\rho/D(z)\right)^2}$, where $D(z)$ is the 
$z$-dependent diameter of the truncated cone-shaped QD. 

The non-secular part of $\oH_{\hbox{\small hf}}$, which is responsible for the spin 
flip-flops as a direct process becomes energetically impossible due to the presence 
of the external magnetic field and the large energy mismatch between nuclear and 
electronic Zeeman energies. On the other hand in the noncollinear case of major EFG 
principal axis being tilted from the growth axis the secular part becomes much more 
interesting.
When we express $\oI_z$ operator in EFG-coordinate components we observe that this allows 
a nuclear spin transition {\em without} changing the spin orientation of the electron.
There is still an energy cost for the nuclear spin flip but this is much less compared 
to non-secular term being on the order of a few neV, 
which is negligible compared to the spontaneous lifetime broadening of $\Delta_l\sim$1~$\mu$eV 
of the exciton state in the optically created electron spin pumping configuration.\cite{borri01} 
Therefore, the rate of noncollinear secular hyperfine interaction (NCSHFI) 
can become significant. For a transition from state $|i\rangle$ to $|j\rangle$, this 
rate is given by
\begin{eqnarray}
\label{Eq-ncshfi}
W^{\hbox{\small NCSHF}}_{ji} & = & f_e
\left[ A_{\hbox{\small hf}} |\psi(\mathbf{R})|^2\right]^2
\left|\langle j|\oI_z|i\rangle \right|^2 
\nonumber\\ & & \times
\frac{2\Delta_l/\hbar}{\left(E_j-E_i\right)^2+\Delta_l^2}.
\end{eqnarray}

\section{Results}
\subsection{Test structures}
There exists a plethora of different realizations for the compositional and structural profiles 
of the self-assembled QDs; 
for a very recent experimental review, see, Ref.~\onlinecite{biasol11}, and specifically for 
the InAs QDs, see, Ref.~\onlinecite{koenraad-chapter}. Guided by the samples used in recent 
nuclear spin experiments,\cite{lai06,eble06,tart07,urbaszek07,makhonin08,kroner08,xu09,latta09,
cherbunin09,makhonin09,eble09,krebs10,cherbunin11} we center our discussion around a QD of a 
truncated cone shape\cite{shape} 
with a base (top) diameter 25~nm (10~nm), placed over a 0.5~nm wetting layer, all fully embedded 
in a GaAs host lattice. The computational supercell contains some 1,800,000 atoms, most of which for
the host matrix region, with the InAs QD region having 16,702 In and 15,432 As atoms. 
In the case of alloy mixing discussions, we randomly replace 
a fraction of the In atoms with Ga atoms. We start from a uniform compressive strain 
in the QD region by setting its lattice constant to that of bulk GaAs. During the energy 
minimization we use periodic boundary conditions while allowing for the computation box 
to shrink and expand, so as to attain a zero pressure on the walls.\cite{parrinello81} 
After this relaxation the InAs QD height settles to 2.93~nm. Our choice for such a 
high-aspect-ratio QD is again guided by the samples used in relevant 
experiments.\cite{lai06,eble06,tart07,urbaszek07,makhonin08,kroner08,xu09,latta09,cherbunin09,
makhonin09,eble09,krebs10,cherbunin11} In order to extract the dependence on the aspect 
ratio and alloy composition of the QDs, different heights and indium mole fractions 
are considered as well.

\subsection{Strain and quadrupolar splitting}
\begin{figure}
\includegraphics[width=8.5 cm]{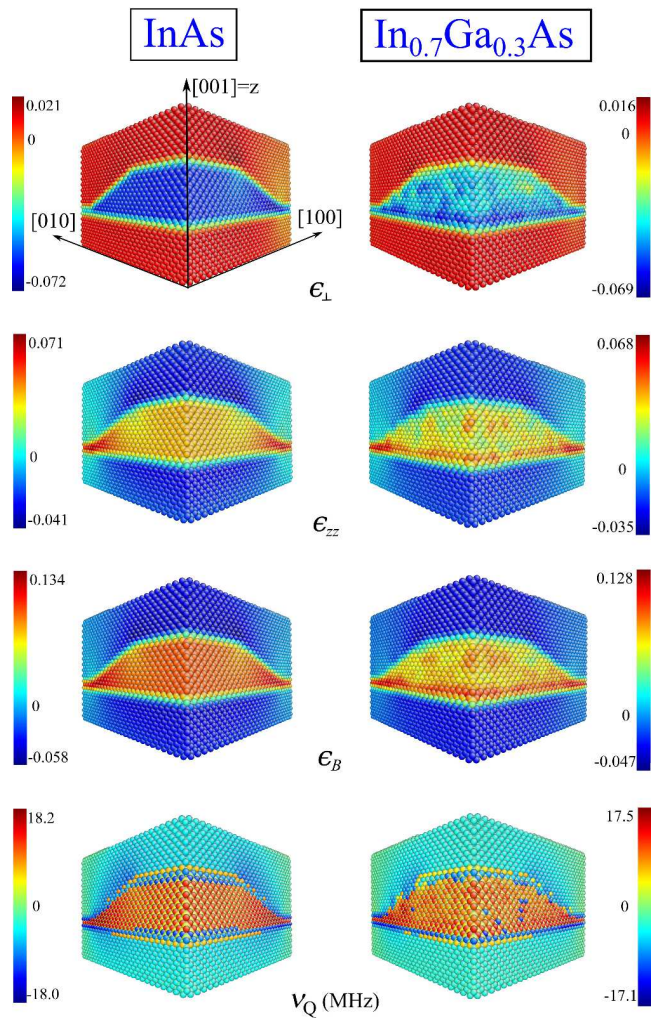}
\caption{\label{fig1} (Color online) Atomistic profiles\cite{atomeye} over (100) and (010) cross 
sections for InAs and In$_{0.7}$Ga$_{0.3}$As QDs. Bigger spheres correspond to In atoms.}
\end{figure}

To have a broad overview, we would like to start with Fig.~\ref{fig1} displaying
the atomistic profiles over (100) and (010) cross sections of InAs and 
In$_{0.7}$Ga$_{0.3}$As QDs. A compressive in-plane strain, 
$\epsilon_\perp\equiv (\epsilon_{xx}+\epsilon_{yy})/2$ is seen 
to be mainly preserved after relaxation whereas for $\epsilon_{zz}$ it is released along 
the growth axis to its environs leaving the QD region with a tensile 
$\epsilon_{zz}$ value.
The opposite signs for the in-plane and growth axis QD strain components is a reflection
of the Poisson effect.\cite{migliorato02} Thus, with the biaxial strain defined as 
$\epsilon_B\equiv\epsilon_{zz}-\epsilon_\perp$, 
a compressive value follows directly 
from these two. It can be observed by comparing left and right panels in 
Fig.~\ref{fig1} that the compositional variation does not lead to
a qualitative change on the strain behavior. In the case of quadrupolar energy parameter, 
$\nu_Q$ the variation among different elements is substantial, while those within 
each element simply follows that of the biaxial strain.

\begin{figure}
\includegraphics[width=8 cm]{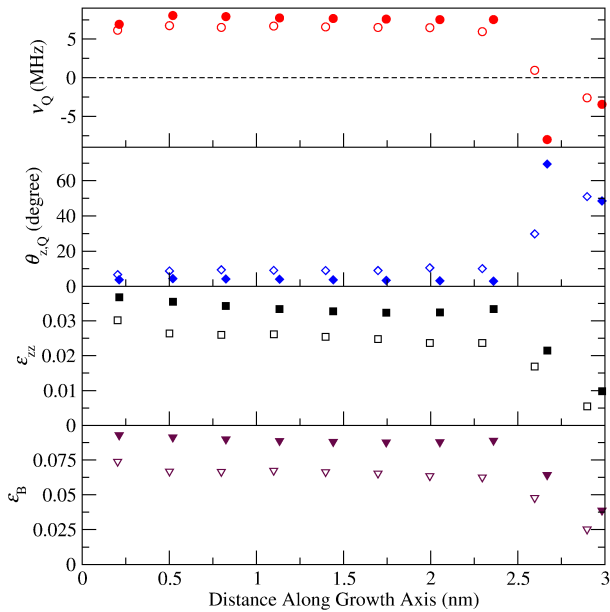}
\caption{\label{fig2} (Color online) Monolayer-averaged axial variations of 
$\epsilon_{B}$ $\epsilon_{zz}$, $\theta_{z,Q}$, and $\nu_Q$ along the 
growth axis starting from the base of the QD without including the wetting 
layer. Filled (hollow) symbols are for InAs (In$_{0.7}$Ga$_{0.3}$As) QD.
The Gaussian envelope square weighting is used for the contribution of atomistic 
quantities within each monolayer.
}
\end{figure}
\begin{table*}
\label{I}
\caption{The mean and standard deviation for certain atomistic quantities of the considered
InAs QD.
$\epsilon_\perp$: in-plane strain perpendicular to growth axis, 
$\epsilon_S$: shear strain, $\epsilon_B$: biaxial strain, 
$\nu_Q$: quadrupolar (lowest) energy splitting, $\eta$: EFG biaxiality parameter,
$\theta_{z,\epsilon}$: polar angle the strain major principal axis makes with the growth axis,
$\theta_{z,Q}$: polar angle the quadrupolar major principal axis makes with the growth axis.
Envelope-square-weighted statistics are given, while the raw (equal-weighted) values are 
quoted in parentheses.}
 \begin{ruledtabular}
 \begin{tabular}{ccccc}
  \multicolumn{1}{c}{} & \multicolumn{2}{c}{Indium} & \multicolumn{2}{c}{Arsenic}\\
\cline{2-3} \cline{4-5}
  & Mean & Standard Deviation & Mean & Standard Deviation \\\hline

$\epsilon_\perp$ & -0.057 (-0.054) & 0.005 (0.012) & -0.056 (-0.055) & 0.004 (0.007) \\
$\epsilon_S$ &  0.004 (0.008) & 0.004 (0.005) & 0.005 (0.010) & 0.004 (0.006) \\
$\epsilon_B$ &  0.090 (0.088) & 0.010 (0.018) & 0.086 (0.084) & 0.013 (0.020) \\
$\nu_Q$ (MHz)& 3.974 (4.074) & 0.427 (0.801) & 9.804 (8.256) & 6.322 (8.955) \\
$\eta$       & 0.042 (0.080) & 0.041 (0.085) & 0.117 (0.255) & 0.198 (0.272) \\
$\theta_{z,\epsilon}$ & 89.3$^{\circ}$ (86.8$^{\circ}$) & 2.3$^{\circ}$ (8.0$^{\circ}$) 
& 88.0$^{\circ}$ (87.6$^{\circ}$) & 1.6$^{\circ}$ (3.4$^{\circ}$) \\
$\theta_{z,Q}$ & 1.7$^{\circ}$ (2.7$^{\circ}$) & 1.6$^{\circ}$ (2.8$^{\circ}$) 
& 12.9$^{\circ}$ (23.7$^{\circ}$) & 24.1$^{\circ}$ (30.6$^{\circ}$) \\
 \end{tabular}
 \end{ruledtabular}
\end{table*}

To support these observations with more quantitative data, in Table~I
the strain and quadrupolar statistics of the InAs QD atoms are summarized.
Since a common way of nuclear spin polarization is via that of an optically oriented
electron spin,\cite{meier84} a relevant measure for the nuclear ensemble would be an 
electron wave function-weighted statistics. Hence, in Table~I and other discussions
we provide equal-weighted and electron envelope-square-weighted statistics for QD nuclear spins,
taking the aforementioned Gaussian form for the latter.
Starting with the strain components, In and As atoms look very similar. The 
exception to this is the shear strain measure that we define as 
$\epsilon_S\equiv|\epsilon_{xy}|+|\epsilon_{yz}|+|\epsilon_{zx}|$, which is 
significantly larger for As. This is caused by those As atoms on the interface 
forming {\em heterobonds} with In and Ga atoms. The shear strain is known to be crucial for 
the piezoelectric field,\cite{schliwa07} whereas in this context it also causes the 
tilting of the quadrupolar axes. Regarding the angular behavior, first of all, the strain major 
principal axis lies almost on the in-plane direction, which is due to the dominance of 
$\epsilon_\perp$.
As a direct consequence of this strain profile, and in particular due to the relation 
$V_{zz}=S_{11}\epsilon_B$, 
the EFG major principal axis is oriented very close to the growth direction. In contrast, 
for the $V_{xx}$ and $V_{yy}$ EFG components, the strain components $\epsilon_{xx}$ 
and $\epsilon_{yy}$ appear in opposite signs and largely cancel. A curious discrepancy 
between In and As occurs for the tilting angle of the major quadrupolar axis from the 
growth direction, $\theta_{z,Q}$: This angle is about eight times larger for As than 
In. The tilting is caused by the shear EFG components, and the discrepancy between In and As 
mainly follows from the $S_{44}$ coefficients, which are $10.0\times 10^{15}$~statcoulomb/cm$^3$ and 
$25.855\times 10^{15}$~statcoulomb/cm$^3$, respectively.\cite{sundfors74} 
In addition, as mentioned before, shear strain is also larger for As, and hence, the larger 
deviation of the As major quadrupolar axis from the growth axis. As an indirect consequence of this, 
As system has twice as large EFG biaxiality, $\eta=(V_{XX}-V_{YY})/V_{ZZ}$ compared to 
In (cf. Table~I).
Another marked difference is about the quadrupolar energy gaps. If we temporarily
assume a uniaxial case ($\eta\rightarrow 0$), the energy difference between 
$m=\pm1/2$ and $m=\pm3/2$ levels becomes $h\nu_{Q}=3e^2qQ/\left[2I(2I-1)\right]$.
This energy is more than twice as large for As ($\sim$8.3~MHz) compared to In ($\sim$4~MHz). The
underlying reason is that In has $I=9/2$ whereas As has 3/2 nuclear spins. The $Q$ values are
however in favor of In, which has $0.860\times 10^{-24}$~cm$^2$ compared to As having 
$0.27\times 10^{-24}$~cm$^2$.\cite{sundfors74} As a matter of fact, if one considers the 
{\em full} nuclear spin manifold energy span under QI, i.e., between $m=\pm 1/2$ and 
$m=\pm I$, In has an extend of $\sim$36~MHz, which is about four times larger than that of As.
If we now focus on the difference between envelope-square- and equal-weighted (in parantheses) 
results, the main deviations are seen to be on the shear strain, EFG tilt angles and biaxiality 
values, which indicate that there comes a substantial contribution from the inclined interface 
regions in the case of equal-weighted statistics. Also note that standard deviation for all 
quantities are higher when the interface region is included 
(i.e., under equal-weighted statistics).

\begin{table*}
\label{II}
\caption{The effect of alloy composition on the Gaussian envelope-square-weighted statistics for 
the same atomistic quantities considered in Table~I.
Three different alloy compositions are compared for a In$_{x}$Ga$_{1-x}$As QD: 
$x=$1, 0.7, and 0.4.}
 \begin{ruledtabular}
 \begin{tabular}{ccccccccc}
  \multicolumn{1}{c}{} & \multicolumn{3}{c}{Indium} & \multicolumn{3}{c}{Arsenic} 
& \multicolumn{2}{c}{Gallium}\\
\cline{2-4} \cline{5-7} \cline{8-9}
  & $x=$1 & $x=$0.7 & $x=$0.4 & $x=$1 & $x=$0.7 & $x=$0.4 & $x=$0.7 & $x=$0.4 \\\hline
$\epsilon_\perp$ & -0.057 & -0.046 & -0.024 & -0.056 & -0.045 & -0.018 & -0.044 & -0.015 \\
$\epsilon_S$ &  0.004 & 0.008 & 0.008 & 0.005 & 0.011 & 0.011 & 0.008 & 0.008 \\
$\epsilon_B$ & 0.090 & 0.067 & 0.034 & 0.086 & 0.065 & 0.031 & 0.068 & 0.033  \\
$\nu_Q$ (MHz)& 3.974 & 3.846 & 2.346 & 9.804 & 8.048 & 1.677 & 4.555 & 2.037  \\
$\eta$       & 0.042 & 0.198 & 0.431 & 0.117 & 0.414 & 0.584 & 0.239 & 0.476 \\
$\theta_{z,\epsilon}$ & 89.3$^{\circ}$ & 87.7$^{\circ}$ & 82.3$^{\circ}$ & 88.0$^{\circ}$ & 86.3$^{\circ}$ & 74.5$^{\circ}$
& 88.3$^{\circ}$ & 70.5$^{\circ}$ \\
$\theta_{z,Q}$ & 1.7$^{\circ}$ & 3.2$^{\circ}$ & 12.4$^{\circ}$ & 12.9$^{\circ}$ & 17.9$^{\circ}$ & 41.5$^{\circ}$
& 5.3$^{\circ}$ & 17.4$^{\circ}$ \\
 \end{tabular}
 \end{ruledtabular}
\end{table*}

Next, in Fig.~\ref{fig2} we return to the effect of alloy mixing and examine the 
variation of monolayer-averaged strain and quadrupolar quantities along the growth axis, 
using Gaussian envelope-square weighting. In all cases the quantities show a rather 
flat profile along the growth axis up until the top two monolayers.
It can be observed that compared to pure InAs case, a substitution of 30\% of 
Ga changes the overall QD strain state in such a way that 
both $\epsilon_{zz}$ and $\epsilon_B$ are reduced, and in corollary, 
so is the energy splitting, $\nu_Q$. On the other hand the tilt angle of the major 
quadrupolar axis from the growth axis, $\theta_{z,Q}$ is larger caused 
by the local random alloy configurations. A large $\theta_{z,Q}$ has a number of 
implications such as enhancing the NCSHFI as will be discussed later.
These observations are further supported quantitatively in Table~II, which compares
the statistics of $\epsilon_\perp$, $\epsilon_S$, $\epsilon_B$, $\nu_Q$, $\eta$, 
$\theta_{z,\epsilon}$, and $\theta_{z,Q}$ for three different alloy compositions.
With the reduction of indium composition from the pure InAs case, the in-plane and biaxial
strain diminishes as well as the quadrupolar energy splitting, whereas, the
increased random alloying enhances the EFG biaxiality and causes substantial deviation 
of the major strain and quadrupolar axes from the growth axis especially for the As and Ga 
nuclei.

\begin{figure}
\includegraphics[width=8 cm]{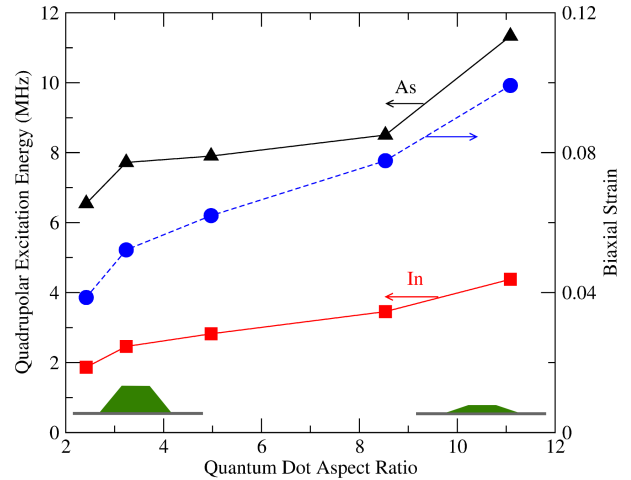}
\caption{\label{fig3} (Color online) Aspect ratio (defined as the base diameter over the 
height) dependence of $\epsilon_{B}$ and $\nu_Q$ for an InAs QD. Base (top) diameter is 
kept at 25~nm (10~nm) while the QD height is varied. 
Envelope-square-weighted values are used.
}
\end{figure}

Finally, we consider the role of aspect ratio of the QD on the strain and quadrupolar state. 
To begin with, if we were 
to have a {\em spherical} QD embedded into a host of a different lattice constant, in the 
continuum approximation we would have only a hydrostatic strain and no biaxial strain 
and no quadrupolar shift. In accordance with this, as seen in 
Fig.~\ref{fig3}, QDs with a large aspect ratio possess a larger biaxial strain causing 
large quadrupolar shifts. 
Note that due to the truncated cone shape, an anisotropy always remains regardless of the
height of the QD. In self-assembled QDs,
while the height is controlled to a very high precision, such as by capping and double 
capping techniques,\cite{koenraad-chapter} this is not the case for the lateral dimension, 
which is essentially determined by the local growth kinetics.\cite{stangl04} Therefore, we can 
expect a significant variance on the quadrupolar shifts from dot to dot within the 
same sample, much like their light emission properties.

\subsection{NMR spectra}
\begin{figure}
\includegraphics[width=8 cm]{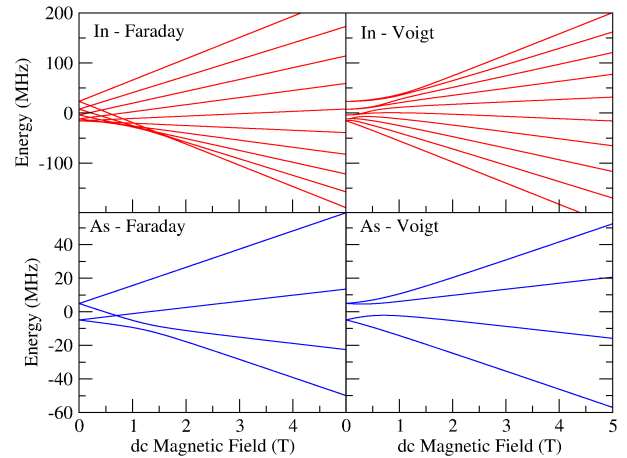}
\caption{\label{fig4} (Color online) QZ spectra of single In and As nuclei for the 
Faraday and Voigt geometries.
}
\end{figure}

\begin{figure}
\includegraphics[width=8 cm]{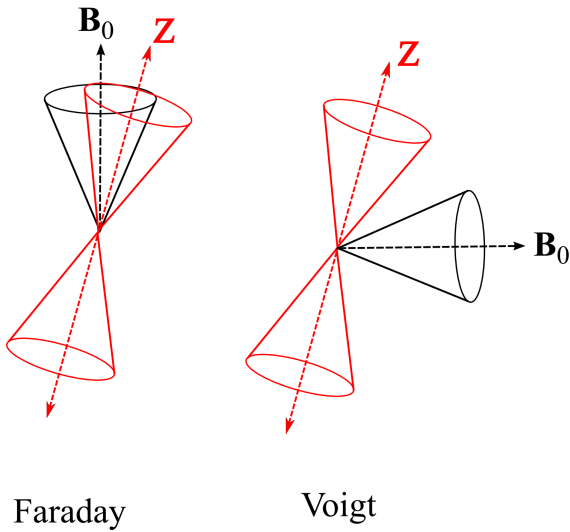}
\caption{\label{fig5} (Color online) Illustration of the magnetic field ($\mathbf{B}_0$), 
and quadrupolar ($Z$) axes for Faraday and Voigt geometries. The single- and bi-cones represent the $|I\rangle$ 
and $|\pm I\rangle$ states for the Zeeman and quadrupolar cases, respectively.
}
\end{figure}

Now, including a dc magnetic field, $\mathbf{B}_0$,
both Faraday and Voigt geometries will be discussed where  
$\mathbf{B}_0$ is parallel and perpendicular to the (optical) growth axis, 
respectively. In Fig.~\ref{fig4} the QZ spectra of a single In and As 
nuclei are given. For the quadrupolar parameters we use those listed 
in Table~I under the envelope-squared
QD statistics. Some of the comparisons regarding the quadrupolar splittings 
between As and In were already made in the previous section. Here, we would 
like to focus on the evolution of the QZ spectra from the quadrupolar- to 
the Zeeman-dominated regime. One can easily realize the marked 
discrepancy between the Faraday and Voigt geometries: in the former, 
as the magnetic field is increased, the quadrupolar-split $m=+I$ state
moves down in energy through a number of band 
crossings followed by a final anticrossing. 
The couplings are localized to the vicinity of these ``crossings''. In the case of 
Voigt geometry however, there are no such band crossings. 
The fundamental difference is illustrated in 
Fig.~\ref{fig5}. Most importantly, QI being a rank-two tensor has a 
{\em bilateral} axis because of which there remains the $\pm m$ degeneracy 
in its spectrum, whereas the magnetic field is vectorial, i.e., {\em unilateral}. 
In the Faraday geometry, as the field 
increases the system loses its bilateral character while slightly rotating to 
align with the magnetic axis. In the Voigt geometry, almost orthogonal setting of the 
two axes causes {\em all} the states to be mixed as the system evolves, as can 
be seen from the corresponding QZ spectra.

We shall now make use of the QZ spectra in interpreting the NMR absorption spectra. 
To excite the nuclear spins a linearly polarized rf magnetic field perpendicular to 
$\mathbf{B}_0$ is introduced. If EFG were both uniaxial (i.e., $\eta=0$) 
and its major principal axis were 
collinear with $\mathbf{B}_0$ then only a perpendicular rf magnetic field would 
cause transitions. In our QDs where these assumptions do not hold, we observed that 
an rf field {\em  parallel} to $\mathbf{B}_0$ also couples to nuclear spins yielding 
a similar spectra. However, their intensity is about two orders of magnitude weaker, 
and for this reason will not be included here.
We assume that initially all nuclear spin states are equally populated 
giving rise to {\em full} spectra, which
can be justified based on the much smaller energy separations among QZ states with 
respect to thermal energy at few Kelvins, as well as due to the presence of dipole-dipole
interactions among the nuclei.\cite{abragam61,slichter78}
As a caveat, we would like to point out that some {\em parts} of the full spectra, to be 
discussed below, can become hindered depending on how the nuclear spin ensemble gets 
prepared under various experimental realizations.

\begin{figure*}
\includegraphics[width=15 cm]{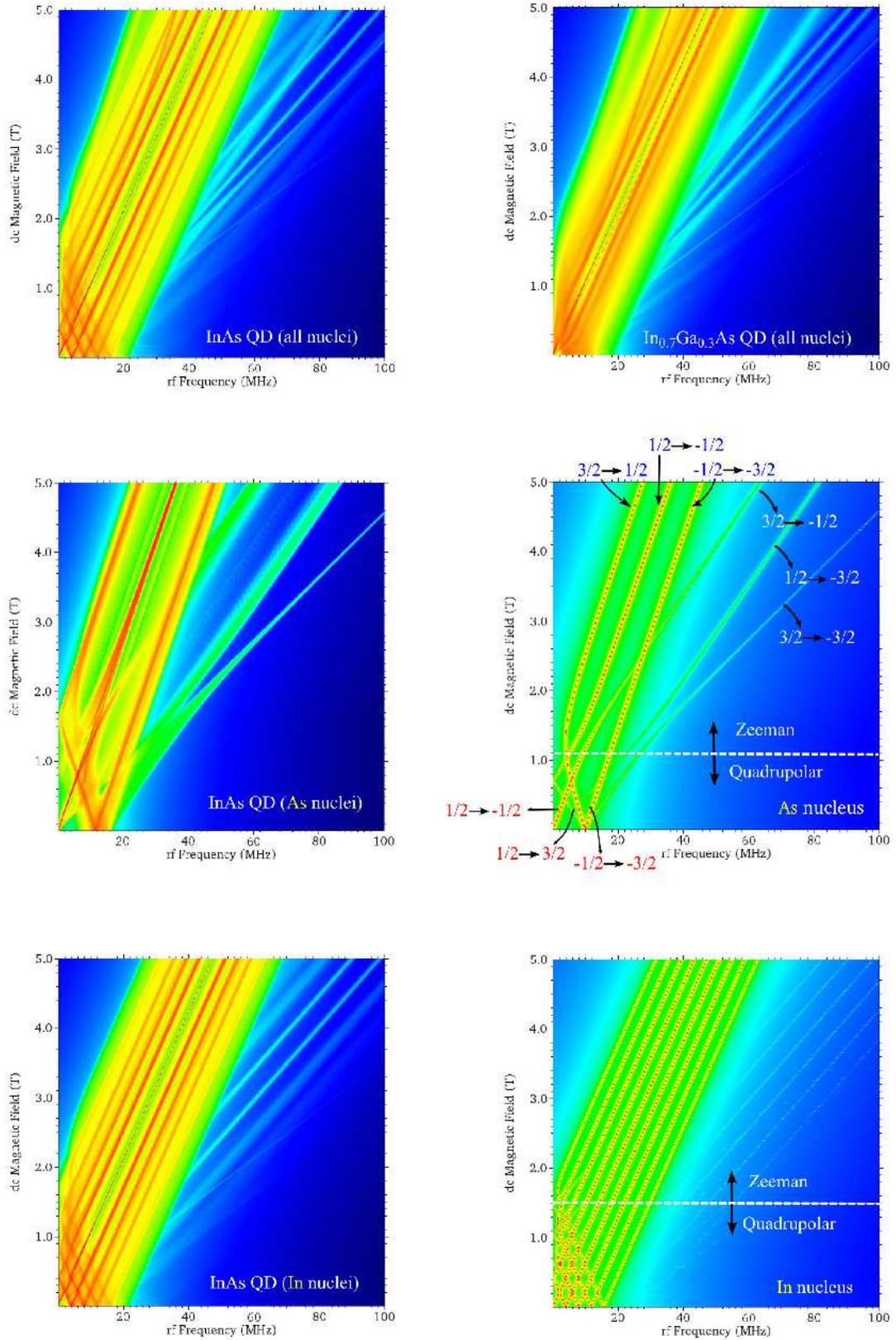}
\caption{\label{fig6} (Color online) The Faraday geometry NMR spectra for InAs and In$_{0.7}$Ga$_{0.3}$As 
QD nuclei (top row), together with their element-resolved contributions, As (center row), 
and In (bottom row) for the InAs QD (left panel, center and bottom rows), contrasted 
with respective single-nucleus spectra (right panel, center and bottom rows).
}
\end{figure*}

\begin{figure*}
\includegraphics[width=15 cm]{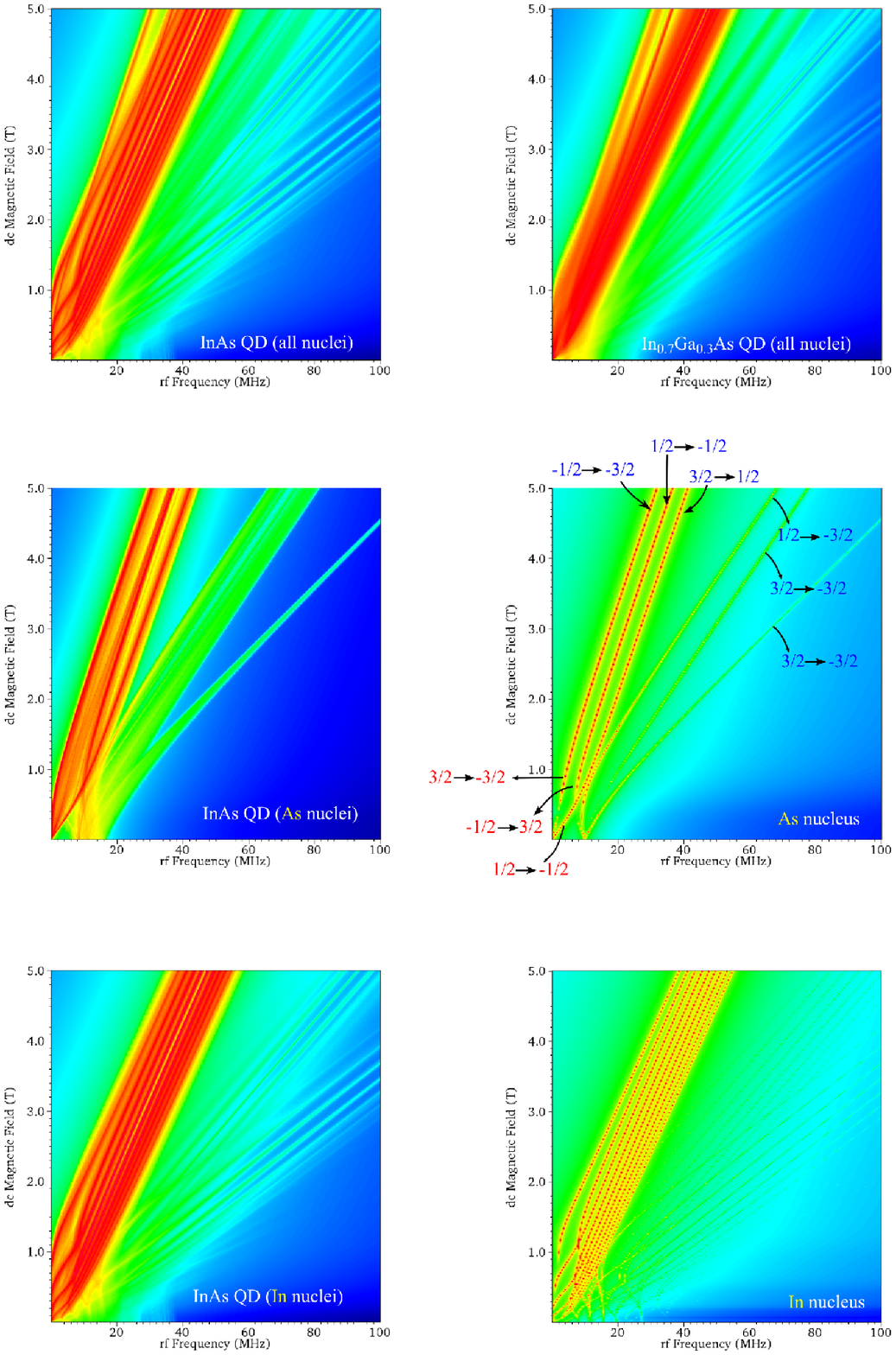}
\caption{\label{fig7} (Color online) Same as Fig.\ref{fig6} but for the Voigt geometry. 
}
\end{figure*}

The NMR spectra of InAs and In$_{0.7}$Ga$_{0.3}$As 
QDs are shown in Figs.~\ref{fig6} and \ref{fig7} for Faraday and Voigt
geometries, respectively. To assist their interpretation we also include those of
the element-resolved and single nucleus spectra as well. As a matter of fact, the 
fingerprints of the single nucleus spectra can be readily identified on the overall QD 
cases. Not observing In nuclei in the Voigt geometry NMR spectra, Flisinski {\em et al.}  
attributed this to the 2.5 times smaller population of each spin state compared to a 
spin-3/2 nucleus.\cite{flisinski10} However, our predictions show that In nuclei must have 
stronger rf absorption intensity compared to As, which stems from the $\gamma I^2$ dependence in the 
rate expression (cf. Appendix); together with the aforementioned population discrepancy this results
in an overall factor of about five in favor of In over As.
For this reason, In nuclei are dominant on the total spectra. As marked in the 
Faraday geometry of single nucleus cases, a borderline between the quadrupolar and 
Zeeman regimes can be introduced, which corresponds to about 1.1~T for 
the As and 1.5~T for the In nucleus. These can be taken as the effective magnetic fields,
$B_Q$ up to which QI is strong. Even though it is not as distinct, 
one can observe that same values also hold for the Voigt geometry.
The transitions among $m$ states are also labeled on the single As nucleus with its 
simpler spectrum, where we use on either side of $B_Q$ the asymptotic pure quadrupolar 
or pure Zeeman basis. For each case individually, the selection rule is\cite{das58} 
$\Delta m=\pm 1$, however, EFG biaxiality or the noncollinear EFG axis with respect 
to $\mathbf{B}_0$ introduce higher-order transitions albeit with much weaker strength. 

The severity of the inhomogeneous broadening can be clearly observed from a comparison 
between the single nucleus and element-resolved spectra. As expected, the higher-order transitions, 
which are forbidden in the absence of QI are highly broadened
because of the atomistic level strain variation over the QD to which they owe their existance.
If we now focus on the overall spectra, we observe that for Faraday geometry, the 
so-called central transition, 1/2 $\rightarrow$ --1/2 is the sharpest among all, 
even in the presence of alloy mixing
in In$_{0.7}$Ga$_{0.3}$As, with the reason being that for this transition QI
has no influence.\cite{mao95} On the other hand for the Voigt geometry {\em all} of single-nucleus 
resonances are broadened due to orthogonality of the quadrupolar axis with respect to 
$\mathbf{B}_0$.

\begin{figure}
\includegraphics[width=8.5 cm]{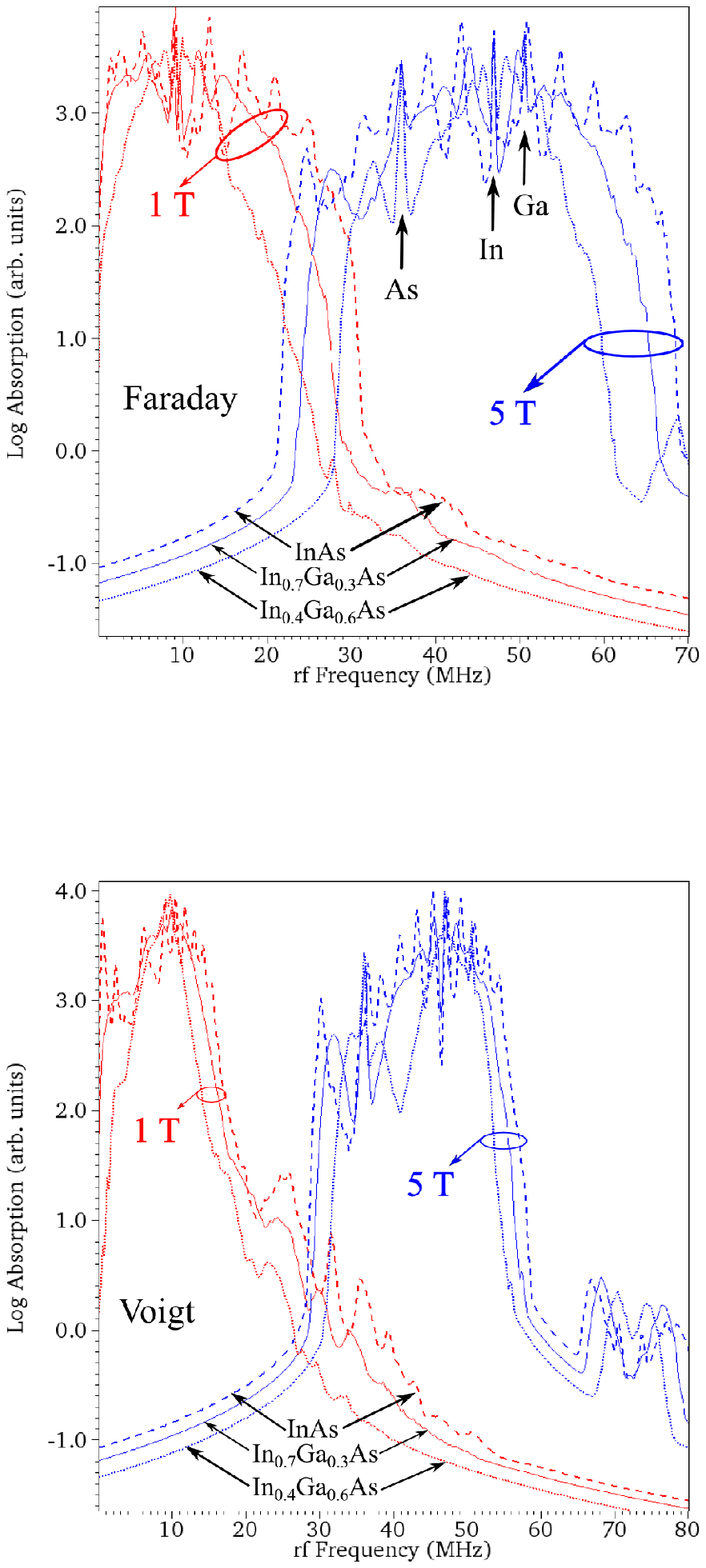}
\caption{\label{fig8} (Color online) Comparison of NMR spectra at 1~T and 5~T for 
InAs (dashed), In$_{0.7}$Ga$_{0.3}$As (solid), and In$_{0.4}$Ga$_{0.6}$As (dotted) QDs. 
Upper (lower) plot is for the Faraday (Voigt) geometry. The vertical arrows in the 
Faraday geometry for 5~T mark the central transitions for In, As, and Ga nuclei.
}
\end{figure}

In Fig.~\ref{fig8} we compare the NMR spectra at 1~T and 5~T for the Faraday and Voigt 
geometries. Note that to gain a broader insight to the compositional effects we also 
include the In$_{0.4}$Ga$_{0.6}$As case. Once again it can be verified that for the Faraday geometry
the central transitions of all QDs coincide with the same sharp resonances, marked by three 
vertical arrows in the 5~T case, corresponding to As (36~MHz), In (47~MHz) and Ga (51~MHz) nuclei. 
Another pivotal observation is that the In$_{0.7}$Ga$_{0.3}$As and In$_{0.4}$Ga$_{0.6}$As 
QDs have progressively narrower overall spectral support with respect to InAs case.
This effect of alloying reminds the motional narrowing in the case of atomic 
gases,\cite{abragam61,slichter78} however, in this case caused by quite a different reason, where
the spin-9/2 manifolds of the In nuclei are partially replaced by the spin-3/2 manifolds of the 
Ga nuclei, with the latter having much narrower span. Therefrom this simply suggests that the NMR 
spectra, especially in the Faraday geometry can be utilized to extract the indium mole fraction, 
which is one of the key unknown material parameters for a specific QD under 
consideration.\cite{biasol11,koenraad-chapter}

\begin{figure}
\includegraphics[width=8.5 cm]{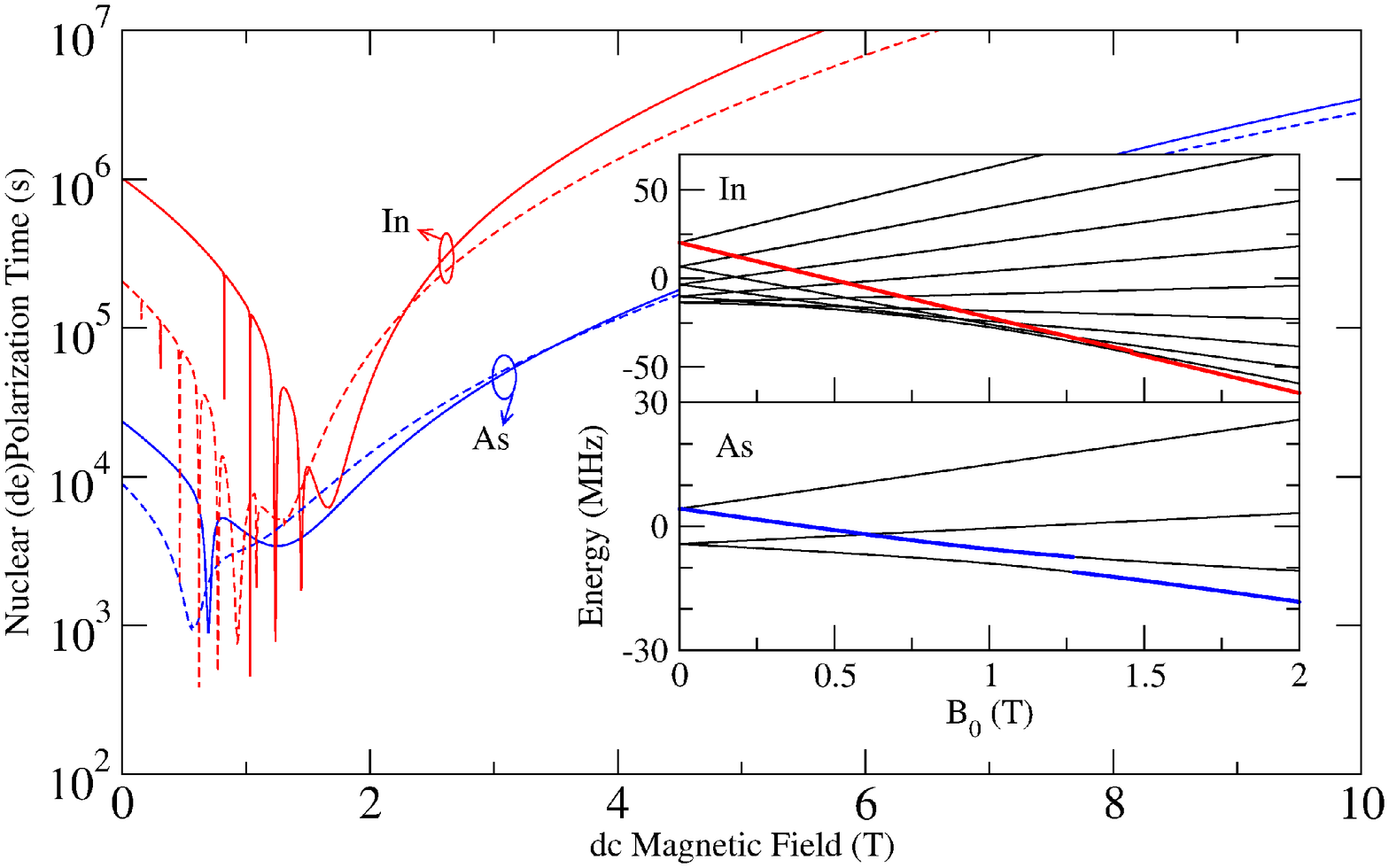}
\caption{\label{fig9} (Color online) Nuclear (de)polarization time of individual In and As nuclei 
due to NCSHFI. QD mean EFG values are used with solid (dashed) lines for InAs 
(In$_{0.7}$Ga$_{0.3}$As). Insets show the evolution of the maximally-aligned state 
(highlighted) among all QZ states as a function of magnetic field. All plots are 
for the Faraday geometry.
}
\end{figure}

\subsection{NCSHFI-mediated nuclear spin depolarization}
Finally, we discuss within the Faraday configuration the NCSHFI-mediated nuclear spin 
depolarization, or for that matter the polarization process as well, as it is 
also governed by the same matrix element (cf. Eq.~\ref{Eq-ncshfi}). 
We assume that nuclear spin is polarized through an electron spin that is
aligned along the growth axis (here, $z$-direction). The presence of the 
quadrupolar field with tilted principal axes introduces a complication, as to which
QZ state is to be ascribed for such a polarized nuclear spin.
For this purpose, we introduce the so-called, {\em maximally overlapping}
state, $|i_{\hbox{\small max}}\rangle$, selected from all QZ states $i\in\{ -I,\dotsc,I\}$ 
that maximizes the overlap, $|\langle m_z|i\rangle|$, where $|m_z\rangle$ denotes a 
free nuclear spin state, which is aligned along the polarizing electron's spin direction.

In the inset of Fig.~\ref{fig9} we see the evolution of the maximally-aligned 
state through a number of band ``crossings'' over the QZ states. 
The total out-transition time from such a state, hence the depolarization time, 
for an individual In or As nucleus under the mean quadrupolar field values (taken from
Table~I) are shown in Fig.~\ref{fig9}. We observe that
the depolarization is enhanced when the initial maximally-aligned state goes through
band crossings with other QZ states, and the broader minimum occurs with the
final band anticrossing. In the case of alloy mixing (dashed lines) larger tilt of the 
EFG axis from the growth direction in general causes rapid depolarization and the 
minimum magnetic field that this occurs also decreases. The minimum NCSHFI-mediated 
depolarization is seen to be on the order of an hour.\cite{cotunnel}
Just as in the NMR spectra, above $B_Q$ level where Zeeman regime takes over, 
the NCSFHI gradually becomes weaker.

\section{Conclusions}
The progress of the optically detected NMR techniques assure that the manipulation 
of a few-number-nuclei will not be a distant 
future.\cite{gammon97,gammon01,makhonin10,makhonin11,flisinski10} 
In line with this prospect, we present the strain and quadrupolar fields that these nuclei 
are exposed to at an atomistic level. First we summarize several structural and compositional
underpinnings: A high aspect ratio enhances the QI, and the interface regions 
introduce biaxiality and the tilting of the major quadrupolar principal 
axis from the growth axis. On the other hand, alloy mixing of gallium into the 
QD reduces both the strain and the quadrupolar energy splitting.
The spectra for Faraday and Voigt geometries are quite distinct from each other.
For the latter, all lines are inhomogeneously broadened
due to the orthogonality of the quadrupolar axis with $\mathbf{B}_0$.
For the former, central transition, 1/2 $\rightarrow$ --1/2 remains sharp even 
in the presence of alloy mixing. Forbidden transitions are also observed,
though highly broadened, arising from the EFG biaxiality and the tilting of 
the quadrupolar axis from the growth direction.
The borderline between the quadrupolar and Zeeman regimes is extracted as 1.5~T 
for In and 1.1~T for As nuclei. At this value the nuclear spin depolarization rate due 
to the noncollinear secular hyperfine interaction with a resident electron in the 
QD gets maximized. In the case of alloy mixing larger tilting of the 
EFG axis from the growth direction causes more rapid depolarization and the 
minimum magnetic field that this occurs also decreases. The shortest NCSHFI-mediated 
depolarization time is seen to be on the order of an hour. As Zeeman regime takes over 
above $B_Q$, this depolarization channel progressively becomes weaker.

\textit{Note added:} After the initial submission of this work a new ODNMR technique
is reported;\cite{chekhovich11} the experimental data for InGaAs QDs at 5.3~T agree extremely well 
with the central transitions marked in Fig.~\ref{fig8}, with the added feature that the 
less common $^{71}$Ga isotope is also resolved, whereas in this work we consider only the
dominant $^{69}$Ga isotope.

\begin{acknowledgments}
We are grateful to Ata\c{c} \.{I}mamo\u{g}lu for suggesting this 
problem and for stimulating discussions. We further thank Martin Kroner, Javier Miguel-Sanchez, 
and P. M. (Paul) Koenraad for discussions regarding the 
experimental connection of this work, and Tahir \c{C}a\u{g}{\i}n and D\"{u}ndar 
Y{\i}lmaz for introducing us to molecular dynamics techniques. Finally, we thank A. I. Tartakovskii
for bringing Ref.~\onlinecite{chekhovich11} to our attention.
\end{acknowledgments}

\appendix
\section*{Appendix: Matrix elements and rf Transition Rates}
In the orthogonal coordinate system as defined by the principal axes of the EFG, the dc 
magnetic field vector $\mathbf{B}_0$ will be in general tilted as governed by the 
spherical polar angles $\theta$, and $\phi$ so that its contribution becomes\cite{das58}
\begin{equation}
\oH_M=-\hbar\Omega\left(\oI_X\sin\theta\cos\phi+\oI_Y\sin\theta\sin\phi+
\oI_Z\cos\theta\right) \nonumber
\end{equation}
where $\Omega=\gamma B_0$. Choosing the angular momentum quantization axis as $Z$, 
and denoting\cite{noteQZ} the free nuclear spin states by $|m\rangle$, we can easily obtain the matrix 
elements of the parts of the full Hamiltonian as
\begin{eqnarray}
\langle m'|\oH_M|m\rangle & = &-\hbar\Omega\left[ m\cos\theta\delta_{m',m}
+\frac{1}{2}(\sin\theta\cos\phi 
\right. \nonumber \\ & & \left. 
\mp i\sin\theta\sin\phi)f_I(\pm m)\delta_{m',m\pm 1}\right],\nonumber \\
\langle m'|\oH_Q|m\rangle & = & A_Q\left\{ \left[ 3m^2-I(I+1)\right ]\delta_{m',m}
\right. \nonumber \\ & & \left. 
+\frac{\eta}{2}f_I(\pm m)f_I(1\pm m)\delta_{m',m\pm 2} \right\} \nonumber,
\end{eqnarray}
where $f_I(m)=f_I(-m-1)=\sqrt{(I-m)(I+m+1)}$, and $A_Q=\frac{e^2qQ}{4I(2I-1)}=\frac{h\nu_Q}{6}$.
Solving for the QZ spectrum essentially yields the expansion coefficients, $C_m^i$ of
the QZ states $|i\rangle$ in terms of free spin states as
\begin{equation}
|i\rangle=\sum_{m=-I}^I C_m^i |m\rangle .
\end{equation}

In the case of an incident rf field, the nuclear spins are excited over 
their established QZ spectrum through the Hamiltonian
\begin{equation}
\oH_{\hbox{\small rf}}=\underbrace{-\hbar\gamma\left[
B_X^{\hbox{\small rf}} \oI_X+
B_Y^{\hbox{\small rf}} \oI_Y+
B_Z^{\hbox{\small rf}} \oI_Z
\right ]}_{\oH'}
\cos\omega_{\hbox{\small rf}} t .
\end{equation}
Hence, based on Fermi's golden rule the rf absorption rate from an initial state $|i\rangle$
to any final state $|j\rangle$ will be
\begin{equation}
W^{\hbox{\small rf}}_{ji}\left( 
\omega_{\hbox{\small rf}}\right)=
\left|\langle j|\oH'|i\rangle \right|^2 \frac{2\Delta/\hbar}{\left(E_j-E_i-\hbar
\omega_{\hbox{\small rf}}\right)^2+\Delta^2},
\end{equation}
where $\Delta$ is the fundamental linewidth of an individual nuclear spin for 
which we take 10~kHz for all the nucleus types in this work.\cite{mao95} The corresponding
matrix element is then given by
\begin{eqnarray}
\langle j|\oH'|i\rangle & = & -\hbar\gamma\sum_{m=-I}^I
B_Z^{ \hbox{\small rf} } \left(C^j_m\right)^* C^i_m m 
\nonumber\\ & &
+B_-^{ \hbox{\small rf} } \left(C^j_{m+1}\right)^* C^i_m f_I(m)
\nonumber\\ & &
+B_+^{\hbox{\small rf}} \left(C^j_{m-1}\right)^* C^i_m f_I(-m) \nonumber
\end{eqnarray}
where 
$B_\pm^{\hbox{\small rf}} =\left( 
B_X^{\hbox{\small rf}} \pm
B_Y^{\hbox{\small rf}} 
\right)/2$.
In the context of NCSHFI having essentially the same matrix element $\langle j|\oI_z|i\rangle$, 
one has to replace the components of $\mathbf{B}^{\hbox{\small rf}}$
in the above expression with the components of the unit vector along the growth axis, $\mathbf{z}$ 
expressed in the EFG coordinate system.


\begin{thebibliography}{99}
\bibitem{note1}In this paper lower case $x$, $y$, and $z$ refer to cubic 
crystallographic axes with $z$ being the growth direction, whereas upper case $X$, $Y$, 
and $Z$ are used for EFG principal axes such that $|V_{XX}|\leq |V_{YY}|\leq |V_{ZZ}|$, 
i.e., the major principal axis chosen along $Z$. Needless to say, shear and biaxial 
components are specified in crystallographic and EFG frames, respectively.
\bibitem{note2}Our notation slightly deviates from Ref.~\onlinecite{shulman57}. 
Throughout this work we make use of tensor strains as opposed to engineering strains.\cite{nye85}
\bibitem{warn}We would like to warn about the recent proliferating 
use\cite{dzhioev07,huang10,flisinski10,chekhovich10} of the expression 
$V_{zz}=S_{11}\epsilon_{zz}$, which is not valid, as the major strain component that 
remains in a QD is $\epsilon_\perp$, whereas $\epsilon_{zz}$ mostly relaxes.
\bibitem{noteQZ}Free nuclear spin states along a general quantization axis will be 
denoted by $|m\rangle$ or $|m'\rangle$, whereas those under both quadrupolar and 
Zeeman fields (i.e., QZ states) will be denoted by $|i\rangle$ or $|j\rangle$.
\bibitem{shape}We also investigated lens and truncated pyramidal shaped QDs and 
observed no qualitative differences. To keep the discussion concise, these results 
are excluded.
\bibitem{cotunnel}However, this should not be
solely attributed NCSHFI, as it requires the presence of co-tunneling and hence a resident 
electron in the dot, without which there is no nuclear spin decay at any 
finite field.\cite{latta11}
\end{thebibliography}
\end{document}